\documentstyle[preprint,eqsecnum,epsfig,prd,aps,floats]{revtex}
\begin{document}

\preprint{\vbox{\hbox{CALT-68-2174}
                \hbox{hep-th/9806226}}}
 
\title{On the M-theory Interpretation of Orientifold Planes}
 
\author{Eric G. Gimon\thanks{email: egimon@theory.caltech.edu}}
 
\address{California Institute of Technology, Pasadena, CA 91125}

\maketitle

\begin{abstract}
We obtain an M--theory interpretation of different IIA orientifold planes by
compactifying them on a circle and use a chain of dualities to get a new IIA
limit of these objects using this circle as the eleventh dimension.  Using 
the combination of the two IIA description, we give an interpretation
for all orientifold four-planes in M-theory, including a mechanism for
freezing M5-branes at singularities.
\end{abstract}

\newpage

\section{Introduction}

	In the past year, M-theory has emerged as useful method for
understanding the non-perturbative dynamics of four dimensional field 
theories\cite{ed1}(for a review and more references, see
Ref.~\cite{review}).  At low energy, higher loop effects of IIA string theory 
can be understood in terms of 11-dimensional supergravity.  A
configuration of NS5-branes, D4-branes, and D6-branes in IIA can
intersect along a common world-volume a produce a variety of possible
four dimensional gauge theories with SU(n) type gauge symmetries.  A
non-perturbative description of these theories' vacua exists in terms
of a single M-theory five-brane wrapping a holomorphic two-cycle in a 
background geometry determined by the D6-branes.  The moduli space of
this curve yields the exact moduli space of the four dimensional field
theory under investigation.  One of the natural extensions of this work was 
to introduce orientifold planes into the IIA picture to get orthogonal
and symplectic groups (see Ref.~\cite{cliff}.  This leads to the
natural question: what do IIA orientifold planes become in the M-theory limit?

	A complete list of possible orientifold O4-planes was recently given 
in the work of Hori~\cite{hori}, along with a natural M-theory interpretation 
for most of them.  One type of symplectic $O4$ plane remained a
puzzle, as it seemed to consist of $M5$-branes mysteriously ``frozen''
at a singularity in a manner reminiscent of Refs.~\cite{karlesp,ednovec}.  
In this paper, we will provide additional information for the
interpretation of $O4$ planes in terms of M-theory by changing the
direction along  which we reduce M-theory to IIA string 
theory.  This can be done by starting with a IIA background containing the $O4$
plane of interest compactified along one of the dimensions of its worldvolume.
Performing a chain of T-S-T dualities along this dimension, we perform the 
so-called ``9-11 flip'', i.e. we exchange the circle along the worldvolume with
the M-circle.  This yields a different IIA description of the same 
M-theory background.  In analogy with the coordinatization of
manifolds, we can think of the two 
IIA backgrounds as charts which make up an atlas for the 
description of the M-theory background.  That is, by putting our ``charts'' 
together, we can reconstruct the full M-theory interpretation for each 
separate type of $O4$-plane.  In particular, we will be able to
uncover how the mysterious ``freezing'' process operates with $O4$ planes.

	This paper will be organized as follows.  We will first review the
T-S-T chain of dualities necessary for our analysis, and demonstrate how this
corresponds to a simple exchange of the circles along the $X^9$ and $X^{10}$ 
directions in the M-theory background.  We will next cover the $O4$
planes which yield orthogonal groups as they can be entirely generated by
the background M-geometry.  We will then add non-dynamical 
five-branes to the mix, providing us with the elements necessary to 
construct $O4$ planes yielding symplectic groups.  We will conclude
with some remarks on what this analysis teaches us about M-theory.

\section{The 9-11 flip}

	In this section we will trace the effects of a chain of T-S-T dualities
on a IIA compactification.  Consider IIA string theory
compactified on a circle, $S^1$, along the $X^9$ direction, with
radius $R$.  We now know that this is in fact M-theory 
compactified on $R^{8,1} \times S^1 \times \bar{S^1}$, where the 
radius $\bar{R}$ controls the string coupling of IIA.  We have:
\begin{eqnarray}\label{acarams}
	&& g^{}_{\rm IIA} = \bar{R} M_S \\	
	&& R_{9} = R \nonumber 
\end{eqnarray}

After we perform a T-duality along the $X_9$ direction, we get a IIB
background with:
\begin{eqnarray}\label{bparams}
	&& g^{}_{\rm IIB} = g^{}_{\rm IIA} {M_S \over R} = 
        {\bar{R} \over R} \\	
	&& R_{9} = \tilde{R} = {1 \over {R M_S^2}} = {1 \over {M_P^3}} 
	{1 \over {R \bar{R}}} \nonumber 
\end{eqnarray}
Now, if we perform an S-duality transformation, we get a new IIB background.
We will label this the IIB' background.  It has :
\begin{eqnarray}\label{b'params}
	&& g^{}_{\rm IIB'} = {R \
over \bar{R}} \\
	&& R_{9} = \tilde{R} = {1 \over {M_P^3}} {1 \over {R \bar{R}}} 
	   \nonumber
\end{eqnarray}
Note that the string length has now been rescaled.

\newcommand{\barA}{\overline{\rm IIA}}

Our final T-duality takes back to a $\barA$ background with
different data:
\begin{eqnarray}\label{a'params}
	&& g^{}_{\barA} = R {M'}_S \\
	&& R_{9} = {1 \over {{M'}_S^2 \tilde{R}}} = \bar{R} \nonumber
\end{eqnarray}
The role of the parameters $R$ and $\bar{R}$ have now been exchanged.  This 
corresponds to exchanging what was the $X^9$ circle, $S^1$, with the M-circle,
 $\bar{S}^1$ (see Refs.~\cite{power,aspinwall}.  We will use the
9-11 flip to trace the descent of objects in 
M-theory within both the IIA and $\barA$ descriptions.

\section{Orientifold four-planes with orthogonal groups}

	The first class of objects that will interest us are orientifold 
four-planes associated with orthogonal gauge groups, i.e. $D4$-branes which 
overlap the plane will carry orthogonal groups.  We will align these such that
they fill the $X^{0}...X^{3}$ and $X^9$ directions.  There are two types of 
 $O4$ planes associated with orthogonal gauge groups.  The first is the 
standard $O4^-$ plane constructed via the quotient 
 ${\mathbf{R}}^{1,9}/\{1,\Omega R_{45678}\}$, where $\Omega$ acts with the 
orthogonal projection and $R_{45678}$ reflects the $X^{4...8}$ directions.  
T-dualizing Type I theory implies that such planes have $-{1 \over 2}$ the 
charge of a $D4$-brane in the bulk. $N$ $D4$-branes from the bulk
which sit on this type of plane will carry an $SO(2N)$ gauge group.  
The second type of $O4$ plane with an
orthogonal $\Omega$ projection is the $O4^0$ plane, which consists of taking 
the $O4^-$ plane we just described and placing a stuck half D4-brane on it (a 
half $D4$-brane comes from just one $D4$-brane in the covering space,
while bulk $D4$-branes come from 2).  We change the $O4$ subscript to
$0$ to indicate that this configuration 
carries no net $D4$-brane charge.  The stuck half $D4$-brane 
adds no new low-energy degrees of freedom to the $O4$ plane, but now a stack of
$N$ $D4$-branes from the bulk sitting on top of the $O4^0$ plane will
carry an $SO(2N+1)$ gauge group.  
Our goal for this section will be to uncover evidence for the nature 
of the $O4^-$ and $O4^0$ plane in M-theory using the 9-11 flip.

\begin{figure}[ht]
\begin{center}
\centerline{\epsfbox{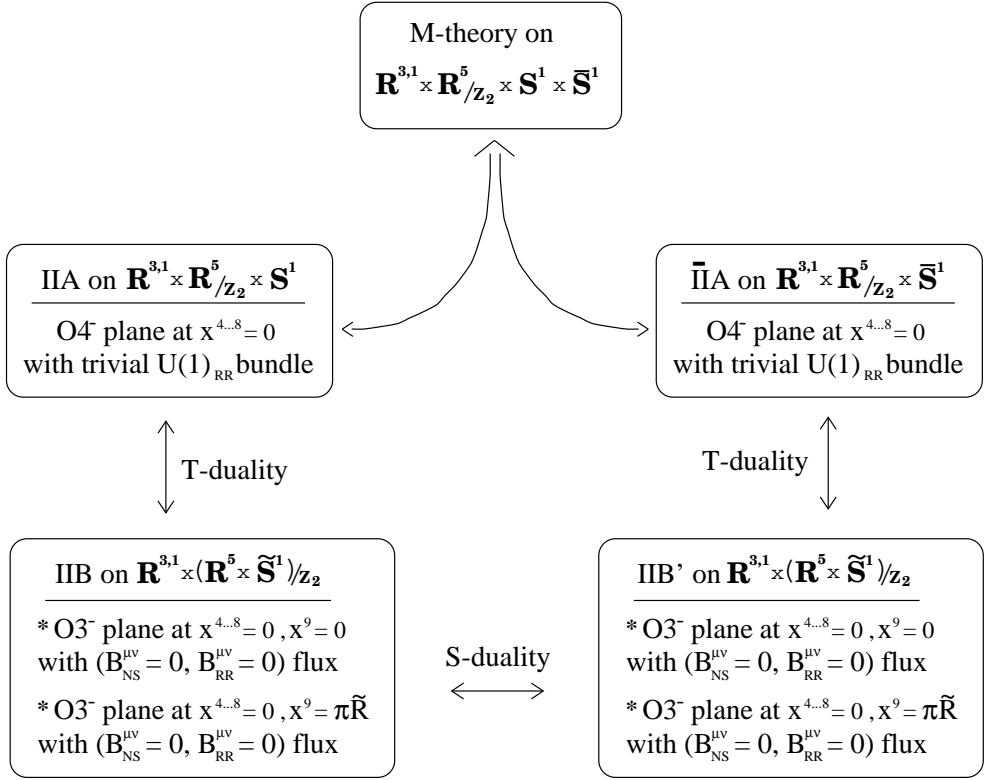}}
\caption{9-11 flip for the $O4^-$ plane}
\label{fig1}
\end{center}
\end{figure}

\subsection{The $O4^-$ plane}

	The $O4^-$ plane is probably the best understood of $O4$ orientifold
planes in M-theory.  We know from Witten's analysis in
Ref.\cite{wittenT5/Z2}(see also Ref.~\cite{mukhi})
 
that the fixed 5-plane in ${\mathbf{R}}^{1,10}/{\mathbf{Z}}_2$, where 
 ${\mathbf{Z}}_2$ reflects the $X^{4..8}$ directions and flips 
the parity of the 
3-form, carries $-{1 \over 2}$ the charge of an $M5$-brane.  Thus this object 
can naturally be identified with the $O4^-$ plane.  If we compactify the $X^9$ 
and $X^{10}$ directions on circles $S^1$ and $\bar{S}^1$ respectively,
we should see no difference between the $O4^-$ plane in the IIA background 
(where $\bar{S}^1$ is the M-circle) and its dual in the $\barA$ 
background.  

	The first step in the T-S-T transformation for the $O4^-$ plane, 
summarized in Fig.~\ref{fig1}, is the T-duality along the $X^9$ direction 
(circle $S^1$).  The $X^9$ direction now has two $O3^-$ planes, one at $X^9=0$
and one at $X^9=\pi\tilde{R}$.  This is a standard result (see for example 
Ref.~\cite{joetasi}), but is also clear from charge conservation.  The $O4^-$ 
plane has D4-brane charge $-{1 \over 2}$, so we expect the T-dual system to 
have D3-brane charge $-{1 \over 2}$.  Since the $O3^-$ planes have D3-brane 
charge $-{1 \over 4}$, everything adds up.  Furthermore, as explained in 
Ref.~\cite{wittenbaryons}, $O3^-$ planes are $O3$ planes with zero NS-NS and 
RR two form fluxes through the ${\mathbf{RP}}^2$ in the ${\mathbf{RP}}^5$ 
surrounding their respective locations in $X^{4...9}$.  We will label them 
 $(0,0)$ planes.  Under S-duality $(0,0)$ planes remain $(0,0)$ planes (see 
Refs.~\cite{wittenbaryons,giveon}).  This ensures that upon further
T-duality 
we now return to an $O4^-$ plane in the $\barA$ background. As expected, 
there is no change under the 9-11 flip.

\begin{figure}[ht]
\begin{center}
\centerline{\epsfbox{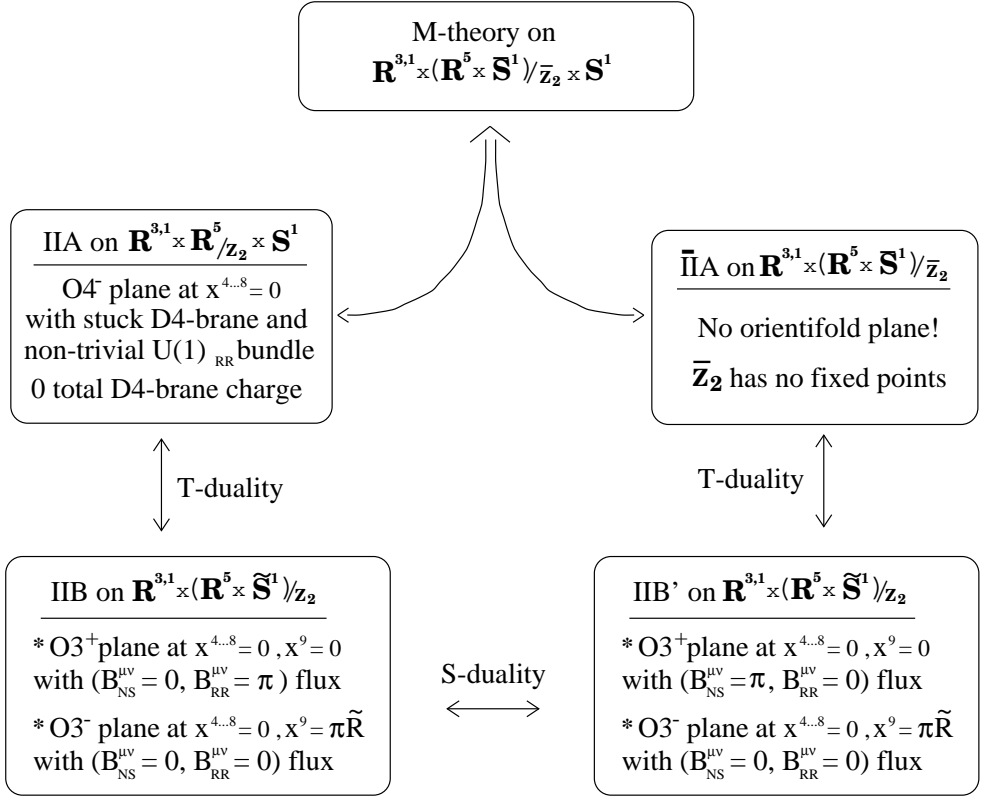}}
\caption{9-11 flip for the $O4^0$ plane}
\label{fig2}
\end{center}
\end{figure}

\subsection{The $O4^0$ plane}

	At this point, we might expect a similarly simple scenario for the 
 $O4^0$ plane.  The naive expectation would be to quotient M-theory on
\begin{equation}\label{naive} 
  {\mathbf{R}}^{1,8}\times S^1 \times \bar{S}^1
\end{equation}
with an $M5$-brane at the origin
using the ${\mathbf{Z}}_2$ that gave us $O4^-$ plane.  In Ref.~\cite{hori}, 
Hori showed that this simple scenario cannot be correct due to flux 
quantization rules.  There is a topological obstruction to having a
{\it single} $M5$-brane at the ${\mathbf{Z}}_2$ fixed plane. It was
suggested that, as an alternative, the $O4^0$
plane might correspond to M-theory on 
\begin{equation}\label{real}
{\mathbf{R}}^{1,4}\times 
 ( {\mathbf{R}}^5 \times \bar{S}^1 ) / {\mathbf \bar{Z}}_2
\end{equation}
where 
 ${\mathbf{\bar{Z}}}_2$ reflects all five directions of ${\mathbf R}^5$
but acts as a translation by $\pi \bar{R}$ on $\bar{S}^1$.  This M\"obius
bundle is everywhere smooth, so doesn't generate any M5-brane charge.  Upon
reduction along $\bar{S}^1$ to IIA, the translation action along $S^1$ 
implies that in the $O4^0$ background, $\Omega R_{56789}$ also flips the
sign of the $U(1)_{RR}$ gauge field~\footnote{Because the $U(1)_{RR}$
gauge field comes from Kaluza-Klein reduction on $\bar{S}^1$, a
translation by $\bar{R} \theta$ along $\bar{S}^1$ acts like
multiplication by the phase $exp(i\theta)$ on the $U(1)_{RR}$ gauge bundle.}.  

	So far, the M\"obius bundle makes an interesting proposal, yet there is
no obvious stringy reason why the presence of a half $D4$-brane on top of
an $O4^-$ plane should necessarily correlate with $\Omega R_{56789}$ acting 
non-trivially on the $U(1)_{RR}$ gauge bundle.  We will use the 9-11 flip to
see that this effect must be there (see Fig.~\ref{fig2}.  The non-trivial
nature of the $U(1)_{RR}$ gauge field first becomes obvious when we perform
the first T-duality of the 9-11 flip.

	Consider a curve, $\hat{C}$ in the $O4^0$ covering space (with the 
 $X^9$ direction now compactified as the circle $S^1$) going from a point to 
its mirror image point under the transformation $\Omega R_{56789}$.  
After we gauge this transformation, we have a closed curve, $C$ in the $O4^0$
background with the property that as we go once around, string states
flip their orientation (due to $\Omega$) and the $U(1)_{RR}$ bundle
flips sign.  Thus we have: 
\begin{equation}\label{flipsign}
e^{i I} \equiv e^{i \int_{C}A^{RR}_{\mu} dx^{\mu}} = -1
\end{equation}
How does this property manifest itself after T-duality? After T-duality the
gauge field integral, $I$, above becomes the integral of $B^{RR}_{\mu\nu}$
along the surface $\hat{C} \times \tilde{S}^1$ 
in $({\mathbf{R}}^5 \times \tilde{S}^1 )/{\mathbf{Z}}_2$ .  We expect
this integral, $I$, to equal $\pi$.  

	From stringy considerations, we know that after T-duality the
 $O4^0$ plane, made up of an $O4^-$ plane and a stuck half D4-brane, will turn
into an $O3^-$ plane at $X^9 = 0$ with a stuck half D3-brane along with a
plain $O3^-$ plane at $X^9 = \pi\tilde{R}$ (the half D3-brane can also be
positioned at $X^9 = \pi\tilde{R}$ with a judicious choice of Wilson lines).
The $O3^-$ plane has D3-brane charge $-{1 \over 4}$, so placing a half D3-brane
on it gives a total charge of $+{1 \over 4}$.  This bound system has total $RR$
two-form flux $\pi$ through any ${\mathbf RP}^2$ inside the ${\mathbf{RP}}^5$ 
surrounding it in $X^{4...9}$ (see Ref.~\cite{wittenbaryons}.  We will refer to
it as an $O3^+$ plane of type $(0,1)$.  $I$ will now be the total flux 
(mod $2\pi$) from the $O3^-$ $(0,0)$ plane and the $O3^+$ $(0,1)$ plane, which
clearly adds up to $\pi$.  Thus we seem to have identified the correct 
originating M-theory configuration for describing the $O4^0$ plane.  

	Continuing along the sequence of dualities which make up the 9-11 flip,
we can gather further evidence for our M-theory picture of the $O4^0$ plane.
After S-duality, the $(0,0)$ and $(0,1)$ $O3$ planes transform into $(0,0)$ and
 $(1,0)$ $O3$ planes respectively~\cite{giveon,wittenbaryons}.  T-dualizing 
this new 
configuration might appear to be difficult, as we are dealing with two 
different type of $O3$ planes.  Fortunately, a very similar type situation,
featuring an $O8^+$ and an $O8^-$ plane was analyzed in Ref.~\cite{ednovec}.
There we start with two $O8$ planes describing the IIA background 
 ${\mathbf{R}}^{1,8}\times \tilde{S}^1/{\mathbf{Z}}_2$ where $\tilde{S}^1$ is
a circle of radius $\tilde{R}$.  After T-duality along $\tilde{S}^1$, a new
type I theory with no vector structure was found with radius 
 ${1 \over 2\tilde{R}}$.  In this theory, however, only {\it even} windings 
are allowed in the cylinder amplitude, and only {\it odd} windings are allowed
in the Klein bottle amplitude.  A simpler way to understand this new theory is
to think of it as the gauging of the IIB theory on 
 ${\mathbf{R}}^{1,8} \times S^1$, $S^1$ with radius 
 $\bar{R} = {1 \over \tilde{R}}$, 
by the operation $\Omega \bar{\mathbf{Z}}_2$ where $\bar{\mathbf{Z}}_2$ acts 
as a translation by $\pi \bar{R}$.

	It is now simple to extend the situation with $O8$ planes of opposite 
charges to the similar case of the $(0,0)$ and $(1,0)$ $O3$ planes.  After 
T-dualizing the $X^9$ direction, we now get $\bar{IIA}$ on
\begin{equation}\label{altiia} 
{\mathbf{R}}^{1,3}\times ({\mathbf{R}}^5 \times \bar{S}^1)/(\Omega R_{5678}
\bar{\mathbf{Z}}_2).
\end{equation}
The second factor describes a M\"oebius bundle.  It has no fixed planes (and
therefore no orientifold planes) and is everywhere smooth.  It does, however,
possess a circle along $X^9$ of minimal length, $\bar{R}/2$ at $X^5=...=X^8=0$
which will become important in the next section.  This $\bar{IIA}$ 
compactification is exactly what we expect if we started from 
Eq.\ref{real} with one of the directions of ${\mathbf{R}}^{1,4}$ on 
compactified and then used this $S^1$ as the M-circle.  We can thus be 
confident that we now have the correct interpretation of the $O4^0$ plane in 
M-theory.

\section{Orientifold four-planes with symplectic groups}
	
	The second class of $O4$ planes we must consider for a complete 
treatment are the $O4^+$ (they have $D4$-brane charge $+{1 \over 2}$) which 
give symplectic groups when we place $D4$-branes on them.  In Ref.~\cite{hori},
it was pointed out that there exists in fact two type of $O4^+$ planes, 
differentiated by there action on the IIA $U(1)_{RR}$ gauge bundle.  As was
the case with the $O4^0$ plane, we can allow $\Omega R_{5678}$ to also act
with a phase $\pm 1$ on this gauge bundle.  If we define the same 
 $A^{RR}_{\mu}$ integral, $I$, as in the previous section, it can take 
values $0$ or $\pi$ modulo $2\pi$.

\subsection{The $O4^+$ plane with $I = 0$}

	After T-duality, $I$ will be the total measure of the $RR$ two-form
fluxes from the $O3^+$ planes at $X^9 =0$ and $X^9 = \pi\tilde{R}$.  Both
of these planes must give $USp(2N)$ type gauge groups.  This means that an
$O4^+$ plane with $I=\pi$ transforms into a $(1,0)$ and a $(1,1)$ $O3$ plane, 
while an $O4^+$ plane with $I = 0$ can transform into a pair of $(1,0)$ $O3$
planes or into a pair of $(1,1)$ $03$ planes.  The ambiguity in the $I=0$ case
might seem confusing at first, until one realizes that the two possibilities
are related by the ${\mathbf SL}(2,{\mathbf Z})$ transformation which
takes $\tau$ to $\tau +1$.  This implies that they correspond different values
for the axion scalar.  We will assume, for now, that the $O4^+$ transforms
into a pair $(1,0)$ $O3$ planes, and the transformation into $(1,1)$ planes 
involves some non-trivial conditions on the $U(1)_{RR}$ gauge field along the
$X^9$ direction.

\begin{figure}[ht]
\begin{center}
\centerline{\epsfbox{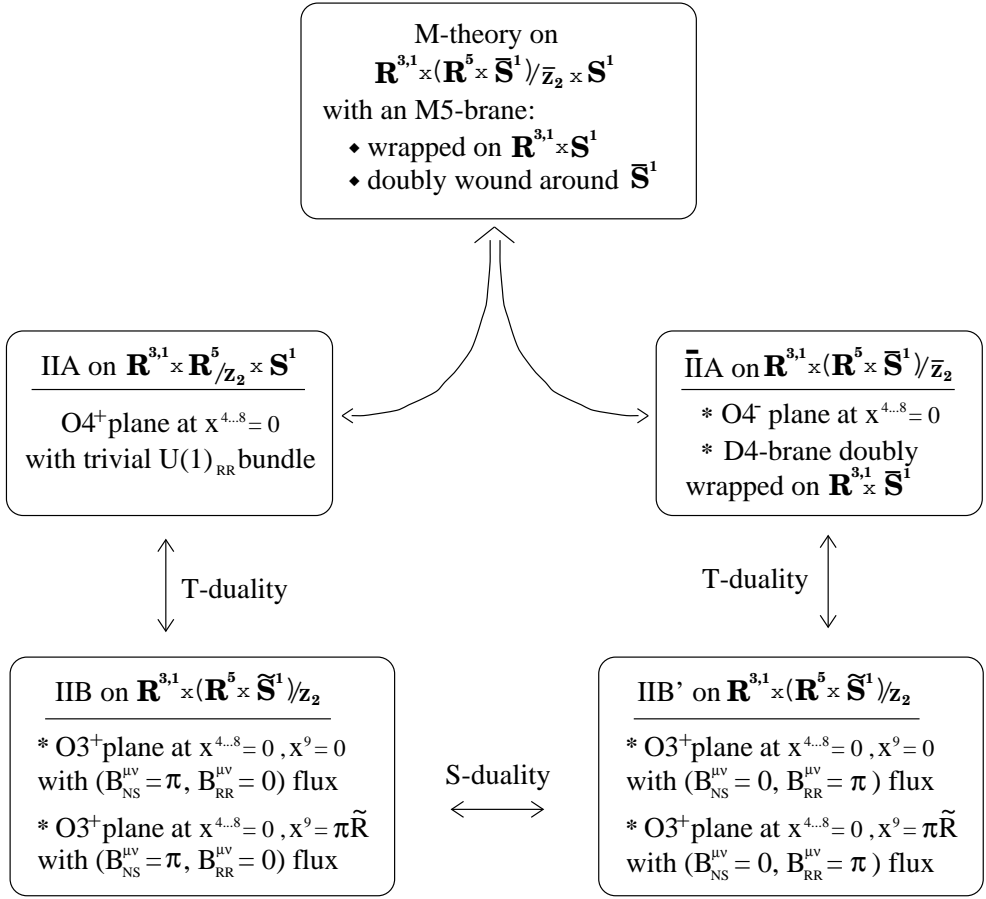}}
\caption{9-11 flip for the $O4^+$ plane with trivial $U(1)_{RR}$ bundle}
\label{fig3}
\end{center}
\end{figure}

	To continue the 9-11 flip for the $O4^+$ case, we next perform an
S-duality (see Fig.\ref{fig3}).  The two $(1,0)$ $O3^+$ planes now become
 $(0,1)$ $O3^+$ planes.  We can think of these two planes as each consisting
of an $O3^-$ plane with a stuck half $D3$-brane.  The final T-duality thus 
yields an $O4^-$ plane with one full $D4$-brane on top of it.  This $D4$-brane,
which normally would be free to move of the $O4^-$ plane, has a special 
 $O(2)$ Wilson line along the $X^9$ direction off the form:
\begin{equation}
\label{wline}
\left( \begin{array}{cc}
	0 & 1 \\
	1 & 0
       \end{array}
\right)
\end{equation}
modulo a gauge transformation.  This implies that open string endpoints on
this $D4$-brane live on a double cover.  Since open strings quantize the
oscillations and zero modes of the $D4$-brane, we know that the $D4$-brane
is wrapped {\it twice} around the circle $\bar{S}^1$. 

	The interesting feature here is that the orientifold projection
prevents the doubly wrapped $D4$-brane from becoming two singly wrapped 
 $D4$-branes.  Before the orientifold projection, the gauge group on the
 $D4$-brane (which is overlapping with its image) is $U(2)$.  In that
gauge group, the Wilson line in Eq.\ref{wline} is continuously connected to
the identity, so there is the possibility of separating the $D4$-brane into
two $D4$-branes with $U(1)$ gauge groups.  While the orientifold projection
is possible once the $D4$-branes are separated (they need only be placed at 
mirror image positions), it does turn $U(2)$ into $O(2)$.  The Wilson 
line in Eq.\ref{wline} is then disconnected from the identity, and the 
separation process is no longer possible.

	Putting together our information from the IIA and $\barA$
backgrounds and decompactifying $S^1$, we see that the $O4^+$ plane with the 
trivial $U(1)_{RR}$ gauge bundle can be thought of as M-theory on
\begin{equation}\label{sp} 
{\mathbf{R}}^{1,4} \times {\mathbf{R}}^5/{\mathbf{Z}}_2 \times \bar{S}^1
\end{equation} 
with an $M5$-brane along ${\mathbf R}^{1,4}$ and doubly wrapped 
around $\bar{S}^1$.  Because the M5-brane is doubly wrapped, we expect that
it will have low-energy excitations with mass of order ${1 \over 2\bar{R}}$.  
We can see this excitation in the IIA picture of the $O4^+$ plane as the
half $D0$-brane stuck on the orientifold plane.   It will have half the mass
of a regular $D0$-brane:
\begin{equation}\label{d0mass}
{1 \over 2}{M_S \over g^{}_{\rm IIA}} = {1 \over 2\bar{R}}.
\end{equation}
The half $D0$-brane should also couple to the $U(1)_RR$ gauge field.

	Finally, if we return to the $O4^+$ plane with $I=0$ wrapped again
on the circle $S^1$, we still need to explain what is necessary for T-dualizing
to a configuration with two $(1,1)$ $O3^+$ plane.  This IIB configuration
is S-dual to itself.  This means that the original $O4^+$ plane must
correspond to an M-theory background like the one in Eq.\ref{sp}, only where 
the $M5$-brane wraps {\it both} $S^1$ and $\bar{S}^1$.  The IIB configuration
with two $(1,1)$ $O3$ planes is also related to one with two $(1,0)$ $O3$
planes by the ${\mathbf SL}(2,{\mathbf Z})$ transformation which takes the
coupling $\tau$ to $\tau +1$.  In IIA variables, this means that the
 $U(1)_{RR}$ gauge field along $X^9$ becomes ${1 \over R}$.  This means that
 $D0$ branes states will be multiplied by a phase $e^(2\pi)$ as they go 
around $S^1$.  Since the half $D0$-brane states have half the charge, they 
will see a phase $e^(\pi)$ as they go around.  These states
effectively live on a circle of twice the radius $R$, demonstrating
how in the M-theory picture we have an $M5$-brane which is doubly
wrapped on both $S^1$ and $\bar{S}^1$. 

\begin{figure}[ht]
\begin{center}
\centerline{\epsfbox{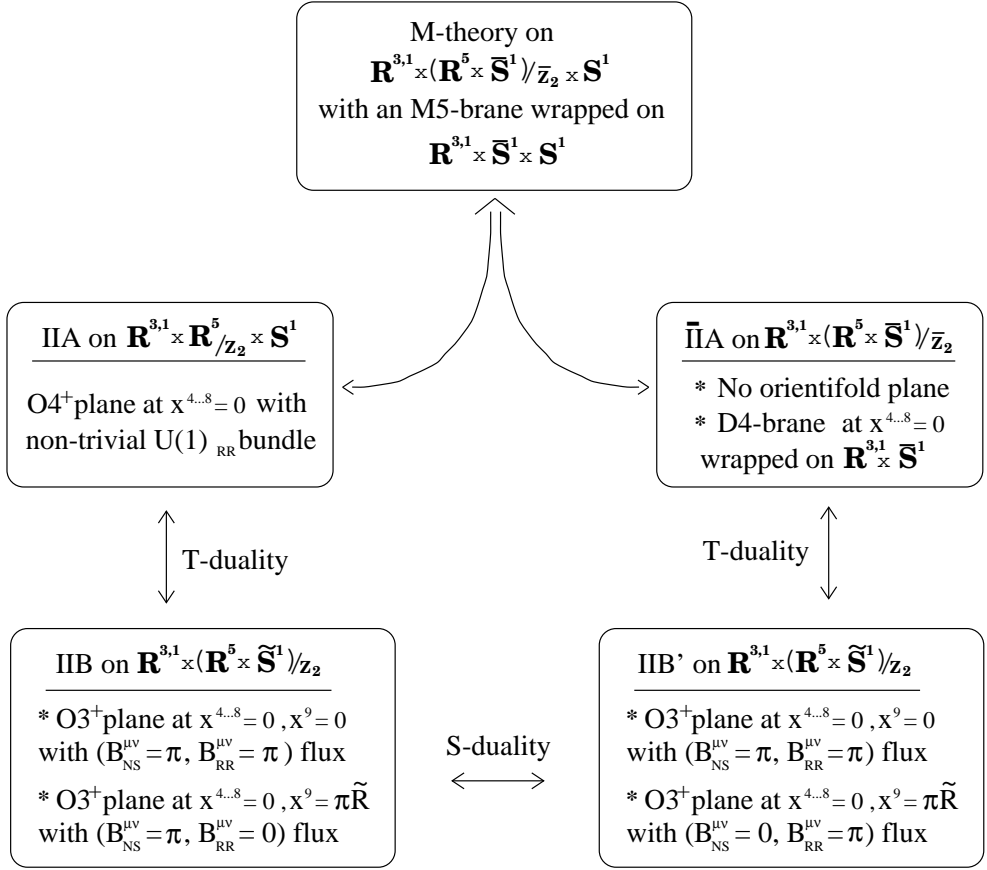}}
\caption{9-11 flip for the $O4^+$ plane with non-trivial $U(1)_{RR}$ bundle}
\label{fig4}
\end{center}
\end{figure}

\subsection{The $O4^+$ plane with $I= \pi$}

	The final $O4$ plane which we would like to consider is the $O4^+$
plane with $I = \pi$.  In Ref.\cite{hori}, it was proposed that in M-theory
this object corresponds to the same background as the $O4^0$ plane, namely
the one Eq.\ref{real}, but with a single $M5-brane$~\footnote{there is no fixed
point set in M-theory, so a single $M5$-brane is allowed in the covering space}
wrapped around $\bar{S}^1$ at $X^4=...=X^8=0$.  Because the 
 $\bar{S}^1/\bar{\mathbf Z}_2$ circle fiber at the origin in 
 $({\mathbf R}^5 \times \bar{S}^1)/\bar{Z}_2$ is the smallest in its
topological class (any deformation of this circle will increase its
overall length), the $M5$-brane can`t move off.  Its lowest energy
excitation will have mass of order ${1 \over 2\bar{R}}$.  In the IIA
background, this corresponds to the half $D0$-brane on the $O4^+$ plane.
We will now use the 9-11 flip, as shown in Fig.~\ref{fig4}, to verify that 
our picture for the $O4^+$ plane with $I = \pi$ is correct.

	The first T-duality in the 9-11 flip gives us two $O3^+$ planes
of type $(1,0)$ and $(1,1)$ respectively.  The difference in $RR$ two-form
fluxes allows us get $I=\pi$.  After S-duality, the $(1,0)$ $O3$ plane becomes
a $(0,1)$ $O3$ plane, while it's $(1,1)$ partner remains the same.  T-duality
now operates just as in the $O4^0$ case, except for two subtleties.   First,
the fact that we have a $(0,1)$ $O3$ plane instead of a $(0,0)$ one means that
the T-dual $\barA$ background will contain a $D4$-brane.  The second
subtlety involves complications with possible non-trivial behavior for the
 $U(1)_{RR}$.  The presence of a $(1,1)$ $O3$ plane opposite the $(0,1)$ $O3$
plane insures that there is no such troubling factor to deal with.  Thus we see
that the 9-11 flip gives us just the result we anticipated, namely a 
 $\barA$ background just as in Eq.~\ref{altiia} with a $D4$-brane wrapped
around the central circle of the M\"obius fibration.

\section{conclusions}

	To summarize, we have shown how $O4$ planes in IIA string
theory arise from two very different types of configuration in
M-theory.  The first class of M-theory backgrounds consists of {\it smooth}
M\"oebius bundles.  They are smooth because we combine a reflection
along five directions with a half-period shift on the M-circle.  The
naked background corresponds to an $O4^-$ plane with a stuck half
$D4$-brane.  We can also dress this background up with an $M5$-brane
wrapped around the circle at the origin to get a special type of $O4^+$
plane.  Both these object will have a non-trivial $U(1)_{RR}$ gauge
bundle which swaps signs under orientation reversal.

	The second class of $O4$ planes involves M-theory on a
singular background, ${\mathbf R}^5/{\mathbf Z}_2 \times ...$.  This
background on it's own corresponds to an $O4^-$ orientifold plane.  We
can, however, wrap a single $M5$-brane at the origin around the
M-circle, so long as it is doubly wrapped.  In IIA, this object
becomes an $O4^+$ plane with a trivial $U(1)_{RR}$ gauge bundle.
For this second class of $O4$ planes, we have been able to learn
something new about M-theory.  Namely, not only does an 
 ${\mathbf R}^5/{\mathbf Z}_2$ singularity carry $M5$-brane charge,
but also a doubly wound M5-brane sitting at this singularity will be
unable to ``unwind'' into two singly wrapped $M5-branes$.

	Looking back, we can identify two main lessons that can be
drawn from this analysis.  First, the 9-11 flip is a useful
method for understanding the M-theory origins of IIA objects.  It
can be used systematically to find these M-theory origins.  If the
M-theory configuration is already known, it can make a useful guide
in understanding what happens under the various dualities that make up
the 9-11 flip.  Another interesting lesson we can draw from this
analysis, is that while there are several type of $O4$ planes in
M-theory, all but one depend on the M-circle compactification to make
sense.  Thus the full uncompactified M-theory has only one
type of orientifold five-plane.  It comes from the 
 ${\mathbf R}^5/{\mathbf Z}_2$ quotient and carries $M5$-charge 
 $-{1 \over 2}$.  If we look at M-theory on $AdS^7 \times S^4$, 
then their is a unique orientation reversing quotient we can
take.  The $(0,2)$ theory of the $2N$ M5-brane that this quotient is
dual to can exhibit both $Sp$ or $SO$ types of gauge symmetries upon
compactification.

\acknowledgments

I would like to thank Clifford Johnson for useful comments and review.
This work was supported in part by the U.S.\ Dept.\ of Energy under Grant no.\
DE-FG03-92-ER~40701.

{\tighten

\end{document}